\documentclass{ifacconf}

\makeatletter
\let\old@ssect\@ssect 
\makeatother

\usepackage{graphicx}
\usepackage{subcaption}
\usepackage{amsmath}
\usepackage{amsfonts}
\usepackage{natbib}        
\usepackage{caption}
\usepackage[dvipsnames]{xcolor}
\usepackage{hyperref}
\usepackage{makecell}

\makeatletter
\def\@ssect#1#2#3#4#5#6{%
  \NR@gettitle{#6}
  \old@ssect{#1}{#2}{#3}{#4}{#5}{#6}
}
\makeatother

\newcommand{\norm}[1]{\left\lVert#1\right\rVert}

\begin{document}
\begin{frontmatter}

\title{\textcolor{black}{In-context learning of state estimators}}


\author[First]{R. Busetto} 
\author[Second]{V. Breschi} 
\author[Third]{M. Forgione}
\author[Third]{D. Piga}
\author[First]{S. Formentin}

\address[First]{Dipartimento di Elettronica, Bioingegneria e Informazione, Politecnico di Milano, Milano, Italy}

\address[Second]{Department of Electrical Engineering, Eindhoven University of Technology, Eindhoven, Netherlands.}

\address[Third]{IDSIA Dalle Molle Institute for Artificial Intelligence USI-SUPSI, Lugano-Viganello, Switzerland.}

\begin{abstract}  
\textcolor{black}{State estimation has a pivotal role in several applications, including but not limited to advanced control design. Especially when dealing with nonlinear systems state estimation is a nontrivial task, often entailing approximations and challenging fine-tuning phases. In this work, we propose to overcome these challenges by formulating an \emph{in-context state-estimation} problem, enabling us to learn a state estimator for a class of (nonlinear) systems abstracting from particular instances of the state seen during training. To this end, we extend an in-context learning framework recently proposed for system identification, showing via a benchmark numerical example that this approach allows us to $(i)$ use training data \emph{directly} for the design of the state estimator, $(ii)$ not requiring extensive fine-tuning procedures, while $(iii)$ achieving superior performance compared to state-of-the-art benchmarks.
}    
\end{abstract}


\end{frontmatter}

\section{Introduction}
\textcolor{black}{State estimation is a core problem in system engineering, as having an estimate of the internal state of a system provides fundamental information on its behavior. The most representative solution to this problem is the Kalman filter, originally conceived for linear systems and has been subsequently extended in different ways to cope with nonlinear systems in both model-based and model-free settings (see \citet{Ghosh2023} for reviews of the two family of approaches). The previous dichotomy becomes relevant when a model for the system whose state one aims at approximating is not known, and one has only data at disposal to design the state estimator.} \textcolor{black}{In this case, one can pursue a (more classical) model-based solution, which however entails the need for an intermediate modeling step. This phase requires additional effort from the designer side, possibly introducing approximation that can deteriorate (or jeopardize) the quality of the retrieved state estimate. On the other hand, model-free techniques do not introduce this kind of approximation error, using data to directly learn the state estimator from data, often at the price of a more intensive data collection (see e.g., \citet{milanese2010direct}) or design phase due to the lack of a model and the structural complexity of the filter to be trained (see, e.g., the techniques reviewed in \citet{Ghosh2023}). Independently from the path one decides to follow, both classes of existing approaches focus on a single system only, thus requiring lengthy re-calibration or re-training of the state estimator when a new system has to be tackled, even if such system is \textquotedblleft similar\textquotedblright \ to the one for which the state estimation problem has already been solved. In turn, this makes the design phase unnecessarily demanding, while not exploiting all available information that has the potential of improving performance (see, e.g., the results attained by exploiting the similarity among systems/control-oriented problems in \citet{chakrabarty2023meta,richards2022control,busetto2023meta}).}\\
\textcolor{black}{To address this limitation of both standard, advanced, and existing model-free solutions to the state estimation problem, in this work we propose to employ the ideas at the core of \emph{in-context learning} \citet{dong2022survey}. The latter is 
a declination of \emph{meta-learning} \citet{vanschoren2018meta}, an approach to obtain a machine learning model that can rapidly adapt to data sampled from a distribution that is --to some extent-- different from the one used for training by leveraging the similarity among different, yet related, problems. Differently from }the most common meta-learning approach, \textit{Model Agnostic Meta Learning} (MAML) \citet{raghu2019rapid}, that entails a bi-level optimization to attain good performance in terms of generalization and adaptation, \textit{in-context learning} consists of training a single --powerful-- (and, thus, likely large) model to successfully deal with new tasks by understanding the underlying context (as in \textit{natural language processing} (NLP)). 
\textcolor{black}{This rationale has been recently extended by \citet{forgione2023context} to tackle \emph{model-free} system identification. There, a Transformer architecture inspired by the one proposed in \citet{vaswani2017attention} is used to directly construct an output predictor without an explicit, task-specific model, which produces one/multi-step-ahead output predictions for a class of dynamical systems when fed with past input/output data.} Following this successful example, \textcolor{black}{in this work we extend this architecture to tackle the state estimation problem by leveraging the structural similarities between this problem and system identification. This allows us to devise}  a model-free \emph{meta-filter}, overcoming at once the typical complexities of nonlinear state estimation. Indeed, the meta-filter is suitable for a whole class of dynamical systems, allowing for an immediate estimate of the state of any system belonging to such class from historical input/output data, without further tuning or optimization. Our numerical results demonstrate the potential of the method for a class of nonlinear systems derived from the process industry, showcasing its superior performance when compared to classical approaches like \textit{Extended Kalman Filters} (EKFs) \citet{awasthi2021survey}, even in the ideal case of full model knowledge. \textcolor{black}{The performance improvement comes nonetheless at the price of increased training and deploying time, the latter due to the non-recursive nature of the in-context-based estimator and the dimension of the meta-filter.} 

\textcolor{black}{The paper is structured as follows. The in-context state estimation problem is formulated in Section~\ref{sec:problem_statement}, where we also compare and contrast the state estimation rationale employed in this work with that of standard and advanced state estimation approaches. In Section~\ref{sec:learning_framework}, we review the structure proposed to tackle the model-free estimation problem addressed in the paper, whose effectiveness is assessed on a benchmark numerical example in Section~\ref{sec:numerical_example}. The paper is ended by drawing some conclusions and indicating directions for future work in Section~\ref{sec:conclusions}.}

\section{Problem formulation}\label{sec:problem_statement}
\textcolor{black}{Let us consider the standard \emph{nonlinear} state estimation framework, i.e., consider a system $S$, whose dynamics is described by the following discrete-time state-pace model:}
\begin{subequations}\label{eq:system}
\begin{align}
    & x_{k+1}=f(x_{k},u_{k})+w_{k},\\
    & y_{k}=g(x_{k},u_{k})+v_{k}.
\end{align}
\end{subequations}
\textcolor{black}{Let $x_{k+1} \in \mathbb{R}^{n_x}$ be the state of the system at time $k \in \mathbb{N}$, $u_{k} \in \mathbb{R}^{n_u}$ be an exogenous (controllable) input, $y_{k} \in \mathbb{R}^{n_y}$ be its measured output, while $w_{k} \in \mathbb{R}^{n_x}$ and $v_{k} \in \mathbb{R}^{pn_y}$ are the process and measurement noise, respectively. Let us assume that the latter are zero mean, white, and uncorrelated.}

\textcolor{black}{Our goal is to design a \emph{filter} $\mathcal{F}_{\phi}$ returning an accurate (in the mean squared error sense) estimate of the system's state at each time step given past input/output data, i.e., 
\begin{equation}\label{eq:filter}
\hat{x}_{k|k}=\mathcal{F}_{\phi}(\mathcal{I}_{k}),~~k=0,1,2,\ldots
\end{equation}
where 
\begin{equation}\label{eq:information_vector}
\mathcal{I}_{k}=\{u_{\kappa},y_{\kappa}\}_{\kappa=0}^{k}
\end{equation}
is the information available at time $k$ to compute the state estimate.}\\
\textcolor{black}{Suppose that the model of the system is unknown and, thus, it should be identified to accomplish this task with classical (model-based) approaches (see, e.g., \citet{awasthi2021survey}). Nonetheless, suppose that we have \emph{prior distributions} for $(i)$ the class $\mathcal{S}$ of dynamical systems to which $S$ in \eqref{eq:system} belongs, $(ii)$ the external inputs, $(iii)$ the process and measurement noises. These priors allow us to generate a (potentially infinite) stream of (finite-dimensional) datasets}
\textcolor{black}{\begin{equation}\label{eq:meta_data}
\mathcal{D}^{(i)}=\left\{u_{k}^{(i)},x_{k}^{\mathrm{o},(i)},y_{k}^{(i)}\right\}_{k=0}^{N-1},~~i=1,2,\ldots,\infty,
\end{equation}}
\textcolor{black}{with $x_{k}^{\mathrm{o},(i)}$ being the $k$-th noiseless state's sample associated with the $i$-th state/input/output sequence. These datasets can be used to learn the filter $\mathcal{F}_{\phi}$ in \eqref{eq:filter} and, thus, solve the state estimation \emph{directly}, bypassing any preliminary (first-principles-based or data-driven) modeling step and lengthy fine-tuning procedures by solving the following \emph{in-context state estimator design} problem:}
\textcolor{black}{
\begin{equation}\label{eq:design_problem1}
\min_{\phi}~~\mathbb{E}_{p(\mathcal{D})}\!\left[\sum_{k=0}^{N-1}\|x_{k}^{\mathrm{o}}-\mathcal{F}_{\phi}(\mathcal{I}_{k})\|^{2}\right],
\end{equation}
with $p(\mathcal{D})$ being the available dataset distribution and $\mathcal{I}_{k}$ defined in \eqref{eq:information_vector}.}
\vspace{-.2cm}
\textcolor{black}{
\subsection{In-context \emph{vs} classical state estimation}
 Traditionally, state estimation is tackled in a model-based fashion. In this setting, when the system $S$ is unknown, the first phase of state estimation generally involves the identification of a (parametric) model for $S$, e.g.,
\begin{subequations}\label{eq:id_model}
    \begin{align}
        &x_{k+1}=f_{\theta}(x_{k},u_{k}),\\
        &\hat{y}_{k}(\theta)=g_{\theta}(x_{k},u_{k}),
    \end{align}
 \end{subequations}
 which can be learned from a set $\{u_{k},y_{k}\}_{k=0}^{N_{\mathrm{id}}-1}$ of (noisy) input/output data as
 \begin{equation}
    \hat{\theta}=\arg\!\!\!\!\!\!\min_{\theta \mbox{ s.t.}\eqref{eq:id_model}}~\frac{1}{N_{\mathrm{id}}}\sum_{k=0}^{N_{\mathrm{id}}-1}\|y_{k}-\hat{y}_{k}(\theta)\|^{2}.
 \end{equation}
The learned model is then used to estimate the state, e.g., by solving 
\begin{subequations}\label{eq:state_est}
    \begin{align}
    &\min_{\psi}~\frac{1}{N}\!\!\sum_{k=0}^{N-1}J(y_{k},u_{k},\hat{x}_{k+1|k},\hat{x}_{k|k}),\\
    &~~ \mbox{s.t. }~ \hat{x}_{k|k}\!=\!{{F}_{\psi}}(\mathcal{I}_{k}),\qquad~~~~~k\!=\!0,\ldots,N\!-\!1,
    \end{align}
    where $F_{\psi}$ is the filter to be designed, the cost is:
    \begin{align}
    \nonumber J(y_{k},u_{k},\hat{x}_{k+1|k},\hat{x}_{k|k})=&\|y_{k}-g_{\hat{\theta}}(\hat{x}_{k|k},u_{k})\|_{R}^{2}+\\
    &+\|\hat{x}_{k+1|k}\!-\!f_{\hat{\theta}}(\hat{x}_{k|k},u_{k})\|_{Q}^{2},
    \end{align}
\end{subequations} 
and $R$, $Q$ are positive definite weights linked to the covariances of the measurement and process noise, respectively.\\  
However, the initial learning step introduces approximations that can be further exacerbated depending on the approach ultimately exploited to design the estimator. This is indeed the case when using \emph{extended Kalman filters} (EKFs) \citet{awasthi2021survey}, where modeling errors are further amplified by the linearization procedure typical of this approach, but also when using sampling-based approaches like \emph{unscented} Kalman filters \citet{Julier2004}, conceived to overcome the limitations of EKF but still inherently model-based. When solving \eqref{eq:design_problem1} the dependence on the model is replaced by that on the distribution $p(\mathcal{D})$, implying that one \textbf{only needs} to know the \textbf{class} $S$ belongs to. In turn, this allows us to design a filter that \emph{adapts} to each plant in the considered family, not requiring re-design for instances of such a systems' class.\\ 
Apart from the possible approximation errors induced by the identification phase, tuning the weights in \eqref{eq:state_est} is not trivial, especially for what concerns the penalty $Q$ on the one-step-ahead prediction. At the same time, fine-tuning $R$ and $Q$ is fundamental for the accuracy of the final estimator \citet{Chen2018WeakIT}, and this procedure has to be carried out every time a new system is considered. In our framework, the calibration of these weights is nonetheless unnecessary. Indeed, by using in-context information, we can design the state estimator by solving a (state) fitting problem, where the only hyper-parameters that we have to tune are those linked to the chosen structure of the filter. Therefore, they do not have to be re-calibrated for every new instance within the class.}

\section{Model-free in-context learning for state estimation }\label{sec:learning_framework}
\textcolor{black}{As it is clear from its definition in \eqref{eq:filter}, the \emph{meta-filter} $\mathcal{F}_\phi(\mathcal{I}_{k})$ represents} a \textbf{unified}, direct mapping between input/output measurements to the states' estimate of any system in a given class. \textcolor{black}{Therefore, the idea behind the model-free, in-context estimator design problem in \eqref{eq:design_problem1} is to exploit the similarities among systems belonging to the same class} to tune it, \textcolor{black}{without the need to specify (and learn) a model for them. Despite being rather intuitive, this idea implies that $\mathcal{F}_\phi(\mathcal{I}_{k})$ has to \textquotedblleft understand\textquotedblright \ the data-generating mechanism and extrapolate how the state is linked to input/output data from the provided context $\mathcal{D}^{(i)}$ (see \eqref{eq:meta_data}), reconstructing the state of the system without returning an explicit model for it nor an explicit representation of the state estimator. To attain this ambitious objective two conditions should be met.} 

\textcolor{black}{First of all, the meta-filter should \emph{ideally} be trained with an infinite amount of input/state/output trajectories from systems sampled from the considered class. From a practical standpoint, this requirement can be relaxed by considering} a sufficiently high number of sampled systems, inputs, and noises \textcolor{black}{at \emph{each iteration} $n_{\mathrm{itr}}$ of \emph{stochastic gradient descent}, thus being able to approximate the loss in \eqref{eq:design_problem1} as follows:}
\begin{equation}
\mathbb{E}_{p(\mathcal{D})}\!\!\left[\sum_{k=0}^{N-1}\!\|x_{k}^{\mathrm{o}}\!-\!\mathcal{F}_{\phi}(\mathcal{I}_{k})\|^{2}\right] \!\!\approx\! 
    \frac{1}{b}\!
    \sum_{i=1}^{b}
    \sum_{k=0}^{N-1}\! \norm{
    x_{k}^{\mathrm{o},(i)} \!\!-\! \mathcal{F}_\phi\!\left(\mathcal{I}_{k}^{(i)}\right)
    }^2\!\!\!,
\end{equation}
\textcolor{black}{where
\begin{equation}
\mathcal{I}_{k}^{(i)}\!=\!\{u_{\kappa}^{(i)},y_{\kappa}^{(i)}\}_{\kappa=0}^{k},~~i\!=\!1,\ldots,b,~~k\!=\!0,\ldots,N-1.
\end{equation}
Increasing the \textquotedblleft batch size\textquotedblright \ $b$ is rather straightforward when a simulator/digital twin for the class of system is available and can be queried an unlimited amount of times. However, getting toward the ideal scenarios where $b \rightarrow \infty$ is less trivial when these conditions are not met, and this aspect is demanded to future works.}\\
\begin{figure}[!tb]
    \centering
    \includegraphics[scale=.8, trim=1cm 18.5cm 11cm 4cm]{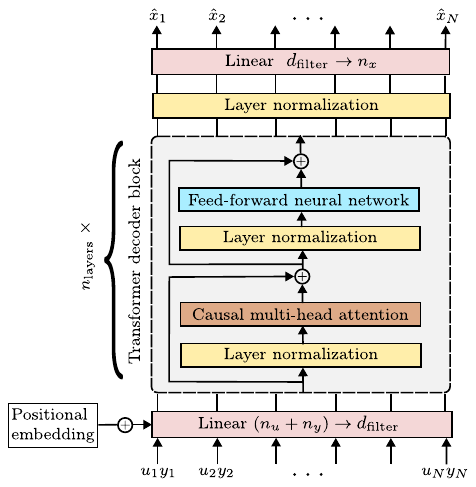}
    \caption{GPT-like decoder-only Transformer for model-free, in-context state estimation.}
    \label{fig:decoder}
\end{figure}
\textcolor{black}{Secondly, the architecture of the filter $\mathcal{F}_{\phi}(\mathcal{I}_{k})$ should be \emph{powerful enough} to generalize over the whole class of systems, thus encompassing their variability. To this end, we adopt the same strategy of \citet{forgione2023context}, adapting the \emph{decoder-only Transformer} architecture for one-step ahead prediction proposed therein (see \citet{vaswani2017attention}) to tackle the state estimation problem. The resulting filter digests the information vector $\mathcal{I}_{k}$ at each time step for the new system, producing the corresponding state estimates as schematized in \figurename{~\ref{fig:decoder}}.}
\textcolor{black}{
The employed architecture is derived from GPT-2 \citet{radford2019language}, a model for NLP applications, here adapted to account for real-valued trajectories. As in \citet{forgione2023context}, this is done by replacing:}
\begin{itemize}
    \item[$(i)$] the first embedding layer with a linear layer mapping, with inputs in $\mathbb{R}^{n_u + n_y}$ and outputs in $ \mathbb{R}^{d_{\mathrm{filter}}}$, 
    \item[$(ii)$] the final layer with a linear layer, with inputs in  $\mathbb{R}^{d_{\mathrm{filter}}}$ and outputs in $\mathbb{R}^{n_x}$,
\end{itemize}
\textcolor{black}{with $d_{\mathrm{filter}}$ being the tunable number of units in each layer. Despite being structurally similar to the open proposed in \citet{forgione2023context}, this last layer represents the main structural difference between the two Transformers. Indeed, it must map the input coming from the inner layers into an output in the state space and not in the output space, coherently with our goal of reconstructing the state of the system.}\\
\textcolor{black}{Apart from these details, the structure schematized in \figurename{~\ref{fig:decoder}} retains all the features of the GPT-2 Transformer, thus inheriting its hyperparameters, i.e., the number of  layers $n_{\mathrm{layers}}$, the number of heads $n_{\mathrm{heads}}$ to be specified for the causal multi-head attention block and the context window length $n_{\mathrm{ctx}}$. It is worth stressing that these parameters have to be tuned only once, not needing a re-calibration when one wants to estimate the state of new instances of the same class of systems.}
\textcolor{black}{
\begin{rem}[Deploying in-context state estimators]\label{rem:deployment}
At every time step $k$, the trained meta-filter $\mathcal{F}_{\phi}(\mathcal{I}_{k})$ relies on the whole information set to return the state estimate. In turn, this implies that the preliminary estimator proposed in this work is \emph{not recursive}, and that the amount of information digested by the filter increases with time. Ultimately, this feature progressively impacts the computational time required to estimate the state. \textcolor{black}{Nonetheless, in practice one can select a sequence of arbitrary length $n_{\mathrm{ctx}}$, the \textquotedblleft context\textquotedblright\ of the Transformer, and employ samples from 1 to $n_{\mathrm{ctx}}$ for $t \leq n_{\mathrm{ctx}}$, or a sliding window from $t-n_{\mathrm{ctx}}$ to $t$ for $t > n_{\mathrm{ctx}}$.}
\end{rem}}
\begin{rem}[One-step-ahead estimation]
\textcolor{black}{The described architecture can be straightforwardly employed to design another filter
\begin{equation}
\hat{x}_{k+1|k}=\mathcal{F}_{\phi^{+}}(\mathcal{I}_{k}),
\end{equation}
returning the one-step-ahead estimation of the state. Future works will investigate the design of this filter and its possible interplay with the one considered here, to reduce the CPU time required for real-time state estimation.}   
\end{rem}
\section{In-context state estimation of an evaporation process}\label{sec:numerical_example}
\textcolor{black}{To assess the performance and effectiveness of the proposed model-free state estimator, let us consider the evaporation process introduced \citet{amrit2013optimizing}, conceived to remove a volatile component from a solvent, concentrating the solution. The dynamics of this process can be described by the second-order state-space model
\begin{subequations}
\begin{align}
& M \dot{x}_{1}=F_{1}X_{1}-F_{2}x_{2}, \\
& C \dot{x}_{2}=F_{4}-F_{5},\\
& y=x_{2},
\end{align}
\end{subequations}
according to which only the second state is directly measured and which, despite what might seem at first glance, is nonlinear. Indeed, while $X_{1}$ is a constant, $F_{1}$, $F_{2}$, $F_{4}$ and $F_{5}$ are functions of the state and the inputs $u_{1}$ and $u_{2}$ according to the dynamics of the different elements of the evaporation process, namely:
\begin{itemize}
    \item the process liquid-energy balance, governed by
    \begin{subequations}
        \begin{align}
            & T_2= ax_{2} + bx_1 + c,~~~T_3= dx_2+e,\\
            & F_4 = \frac{Q_{100}-F_1 C_p(T_2-T_1)}{\lambda};  
    \end{align}
    \end{subequations}
    \item the heat steam jacket, driven by
    \begin{subequations}
    \begin{align}
    & T_{100} = \varphi u_{1}+\gamma, ~~~Q_{100} = UA_1(T_{100}-T_2),\\
    & UA_1 = h(F_1+F_3),~~F_{100}= {Q_{100}}/{\lambda_s};   
    \end{align}
    \end{subequations}
    \item the condenser, for which
    \begin{equation}
        Q_{200}= \frac{UA_2(T_3 - T_{200})}{1+UA_2/(2C_pu_{2})},~~~F_5 = \frac{Q_{200}}{\lambda},
    \end{equation}
    holds;
    \item and, lastly, the level controller, such that:
    \begin{equation*}
    F_2 = F_1 - F_4.    
    \end{equation*}
\end{itemize}
In turn, all these equations feature a set of coefficients, whose nominal values are reported in \tablename{~\ref{tab:process_params}}. These values are uniformly perturbed at random of up to $20\%$ of their nominal values to construct the instances of the system class used to construct the meta dataset in \eqref{eq:meta_data} and to generate any new instance of the evaporation process. All instances of the class are affected by Gaussian-distributed process and measurement noises, characterized by
\begin{equation}\label{eq:noises}
w \sim \mathcal{N}(0,\mathrm{diag}(0.5,0.5)),~~~v \sim \mathcal{N}(0,2),
\end{equation}
that are white and mutually independent. Output data are always collected with a sampling time of $T_{s}=1$~[s].}

\begin{figure}[!tb]
    \centering
    \includegraphics[scale=.6]{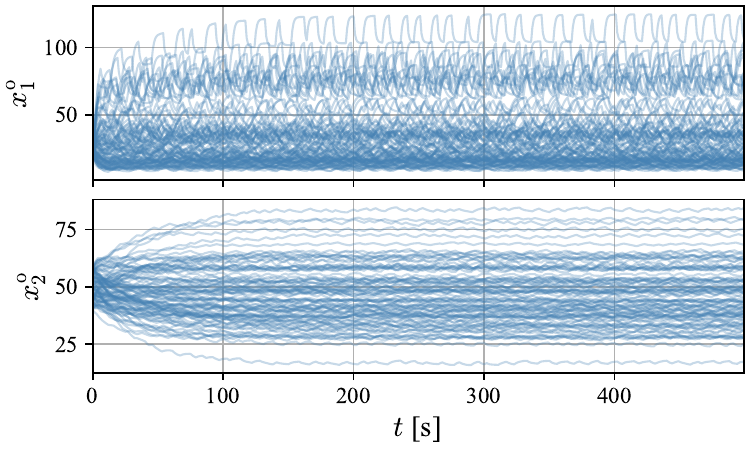}\vspace{-.2cm}
    \caption{\textcolor{black}{Ground-truth state from $100$ unseen instances of the evaporation process used for meta-filter testing.}}
    \label{fig:meta-test-set}
\end{figure}

\begin{figure}[!tb]
\centering
\begin{subfigure}[h]{\linewidth}
    \centering
    \includegraphics[scale=.575]{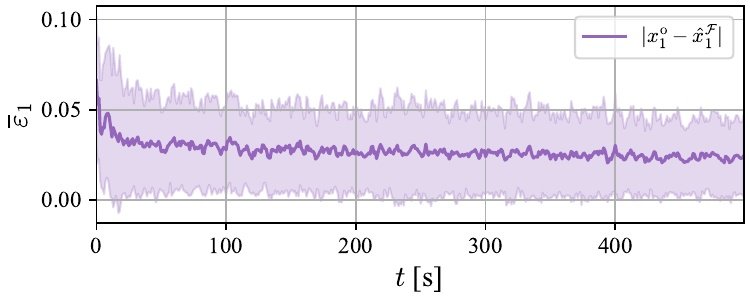}\vspace{-.2cm}
    \caption{Absolute estimation error $\overline{\varepsilon}_{1,t}=|x_{1,t}-\hat{x}_{1,t|t}|$.}
    \label{fig:error_x1_transformer}    
\end{subfigure}\vspace{.1cm}
\begin{subfigure}[t]{\linewidth}
    \centering
    \includegraphics[scale=.575]{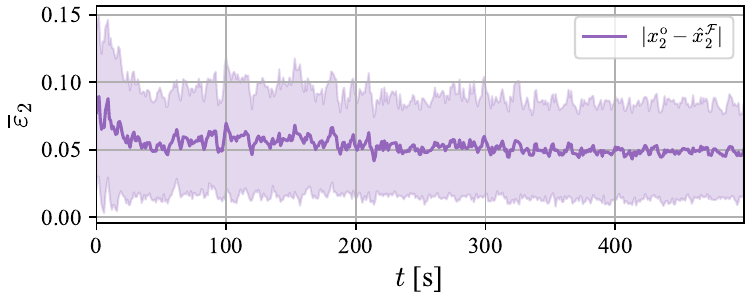}\vspace{-.2cm}
    \caption{Absolute estimation error $\overline{\varepsilon}_{2,t}=|x_{2,t}-\hat{x}_{2,t|t}|$.}
    \label{fig:error_x2_transformer}    
\end{subfigure}\vspace{-.2cm}
\caption{\textcolor{black}{Absolute state estimation error in testing: mean and standard deviation (shaded area) over $100$ instances of the evaporation process unseen in training.}}
\end{figure}

\textcolor{black}{To generate the datasets $\mathcal{D}^{(i)}$ in \eqref{eq:meta_data}, we consider the steady-state state and input values achieved with the evaporation process having the parameters reported in \tablename{~\ref{tab:process_params}}, namely \begin{equation}\label{eq:nominal_ss_values}
    x_s = (25, 49.743), \qquad u_s = (191.713, 215.888).
\end{equation}
Starting from these values, we perturb the initial condition of each instance of the evaporation process as follows:
\begin{align*}
    x_{0,1} &= x_{1,s} ( 1 + 0.2\cdot\mathcal{U}_{[-1,1]})\\
    x_{0,2} &= x_{2,s} ( 1 + 0.2\cdot\mathcal{U}_{[-1,1]}),
\end{align*}
where $\mathcal{U}_{[-1,1]}$ denotes a uniform distribution in the interval $[-1,1]$. Each instance of the process is fed with a perturbed version of the steady-state input $u_s$, constructed as follows:
\begin{align*}
    u_{1,k} &= u_{1,s} +\text{PRBS}_k,\\
    u_{2,k} &= u_{2,s} +  \text{PRBS}_k,
\end{align*}
for $k=0,\ldots,500\cdot T_{s}$, where the PRBS signal assumes values $\pm 20$. These experiments result in noiseless state trajectories like the ones reported in \figurename{~\ref{fig:meta-test-set}}, which represent the ground truth to which we compare the outcome of the learned meta-filter in testing.} 
\begin{table*}[t]
\caption{Nominal model parameters \textcolor{black}{for the evaporation process.}}
\label{tab:process_params}
    \centering
    \begin{tabular}{ccccccccccccccccccc}
        \hline
        $a$ & $b$ & $c$ & $d$ & $e$ & $\varphi$ & $\gamma$ & $h$ &
        $M$ & $C$ & $UA_2$ & $C_p$ & $\lambda$ & $\lambda_s$ & $F_1$ & $X_1$ &
        $F_3$ & $T_1$ & $T_{200}$\\ 
        \hline
        0.5616 & 0.3126 & 48.43 & 0.507 & 55 & 0.1538 & 55 & 0.16 &
        20 & 4 & 6.84 & 0.07 & 38.5 & 36.6 & 10 & 5 &
        50 & 40 & 25\\
        \hline
    \end{tabular}
    \label{tab:my_label}
\end{table*}
\textcolor{black}{
The in-context state estimator $\mathcal{F}_{\phi}$ satisfying \eqref{eq:filter} is trained on an x64-based laptop, equipped with an 8 Core-processor i7-10875H, 16 GB RAM, and an Nvidia RTX 2060 GPU. The proposed architecture is implemented\footnote{The code is adapted from \citet{forgione2023context}, as the structure of the Transformer has not been modified, and it is available at \url{https://github.com/buswayne/sysid-transformers-control}.} within the Pytorch framework \citet{paszke2019pytorch}, using the hyper-parameters reported in \tablename{~\ref{tab:hyperparams}}. This table further reports the time and \emph{root mean square error} (rmse) attained in training, already showing that the filter accurately reconstructs the state in training despite the different instances of the evaporation process used, at the price of a time-intensive training phase. Nonetheless, it is worth pointing out that such an effort has to be carried out offline, thus not comprising the usability of the filter.}\\
\begin{table}[!tb]
    \centering
    \caption{\textcolor{black}{Hyper-parameters, the training time and the resulting rmse.}}
    \label{tab:hyperparams}
    \setlength{\tabcolsep}{3pt}
    \begin{tabular}{ccccccccc}
        \hline
        $n_{\mathrm{param}}$ & $n_{\mathrm{layers}}$ & $n_{\mathrm{heads}}$ & $n_{\mathrm{ctx}}$ & $d_{\mathrm{filter}}$ & $n_{\mathrm{itr}}$ & $b$ & train time & rmse
        \\
        \hline
        2.45 $\cdot 10^{6}$ & 12 & 4 & 500 & 128 & 50 $\cdot 10^{3}$  & 32 & 12 h & 0.06\\
        \hline
    \end{tabular}
\end{table}
\textcolor{black}{By using the meta-filter trained as specified before and using it to reconstruct the $100$ instances of the ground-truth (noiseless) states reported in \figurename{~\ref{fig:meta-test-set}}, we obtain the absolute estimation errors reported in \figurename{\ref{fig:error_x1_transformer}} and \figurename{\ref{fig:error_x2_transformer}} for the states $x_1$ and $x_2$, respectively.} From these results it is possible to appreciate that the average absolute estimation error $\overline{\varepsilon}$ \textcolor{black}{tends to decrease over the considered horizon, becoming relatively small after an initial transient of about $50$ s. This transient is due to the time required for the Transformer to \textquotedblleft understand\textquotedblright \ the new context from the information that is recursively fed to it. A similar effect can be noticed for the standard deviation of the absolute estimation error, which tends to reduce after the initial transient. By looking at \figurename{\ref{fig:error_x1_transformer}} and \figurename{\ref{fig:error_x2_transformer}} it is also clear that the quality of the estimate for the unobserved state $x_{1}$ is better (in terms of absolute estimation error) than the one of the measured state $x_{2}$. Despite this result being quite counter-intuitive, the reason behind this behavior can be easily grasped by looking at \figurename{~\ref{fig:meta-test-set}}. Indeed, from the ground-truth (noiseless) states it is clear that $x_1$ is more responsive to changes in the input, thus making it easier for the meta-filter to grasp changes in context for this state rather than the measured one.
}
\begin{figure}[!tb]
    \centering
    \includegraphics[scale=.575]{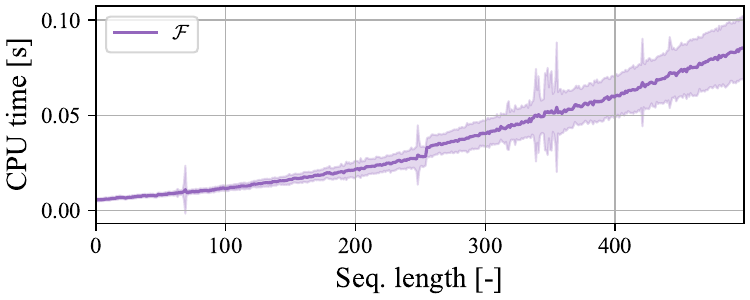}\vspace{-.2cm}
    \caption{\textcolor{black}{CPU time to retrieve estimate in testing: mean and standard deviation (shaded area) over $100$ instances of the evaporation process unseen in training.}}
    \label{fig:computational_time}
\end{figure}
In terms of computational time, \textcolor{black}{the meta-filter behaves as depicted in \figurename{~\ref{fig:computational_time}}, namely the time required to estimate the state increases with time. Nonetheless, such behavior is expected since the Transformer has to digest the entire information vector to extrapolate the estimate of the state.}

\subsection{Comparison with benchmark filters}
\textcolor{black}{We now compare the performance of the meta-filter proposed in this work against the ones attained with two versions of the extended Kalman filter. On the one side, we consider the ideal case in which the model of the system is fully known for each new instance of the evaporation process, referring to the associated EKF as the \emph{Oracle} EKF (O-EKF). On the other hand, we assume that the model is overall known but that one parameter (specifically $UA_2$) is unknown and it has to be estimated from data. The state space model of the evaporation process is thus augmented by a third state, with dynamics
\begin{equation}
    \dot{x}_{3} = 0,
\end{equation}
leading to what we refer to as the \emph{Enlarged-state} EKF (E-EKF). Note that, this choice still implies that we inject the correct prior on the nature of the unknown parameter, thus still putting ourselves in a semi-ideal case. For both EKFs, we fix the weighting matrices of the extended Kalman filter ($Q$ and $R$) to the true values of the covariance matrices of the process and measurement noise (see \eqref{eq:noises}), extending $Q$ to $\mathrm{diag}(0.5,0.5,0)$ in the E-EKF case. Instead, the covariance matrix of the estimation error is fixed to $P_0=\mathrm{diag}(0.1,0.1)$ for O-EKF, and extended to $P_{0}=\mathrm{diag}(0.1,0.1,1)$ for E-EKF. This parameter has been chosen based on an extensive sensitivity analysis but is nonetheless not re-calibrated for new instances of the evaporation process, to show the impact that the lack of fine-tuning has on the performance of EKF. In both cases, the EKF is implemented in Python using the supporting library of \citet{labbe2014kalman} and CasAdi \citet{andersson2019casadi} for model prediction and linearization.}\\
\begin{figure}[!tb]
\centering
\begin{subfigure}[h]{\linewidth}
    \centering
    \includegraphics[scale=.575]{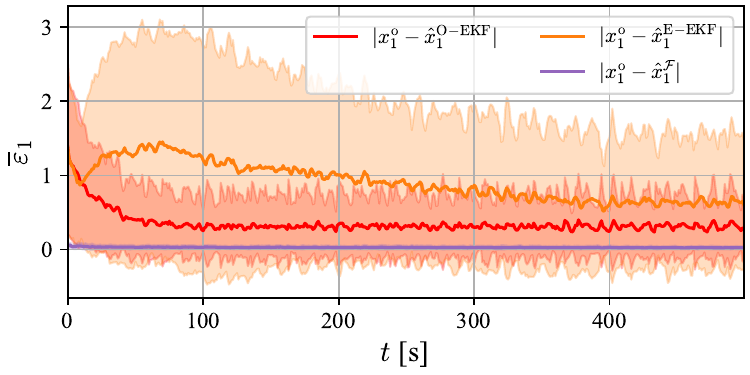}\vspace{-.2cm}
    \caption{Absolute estimation error on $x_1$.}
    \label{fig:lineplot_e1}    
\end{subfigure}\vspace{.1cm}
\begin{subfigure}[t]{\linewidth}
    \centering
    \includegraphics[scale=.575]{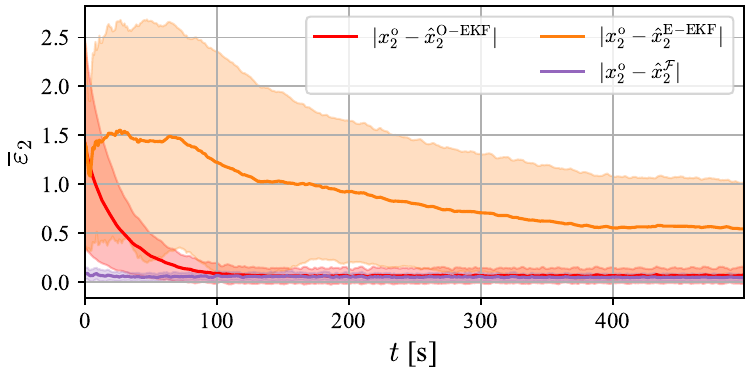}\vspace{-.2cm}
    \caption{Absolute estimation error on $x_2$.}
    \label{fig:lineplot_e2}    
\end{subfigure}\vspace{-.2cm}
\caption{\textcolor{black}{Absolute state estimation error in testing: mean and standard deviation (shaded area) over $100$ instances of the evaporation process unseen in training attained with the meta-filter, O-EKF and E-EKF.}}\label{fig:comparison_errors}
\end{figure}
\begin{figure}[!tb]
    \centering
    \includegraphics[width=\linewidth]{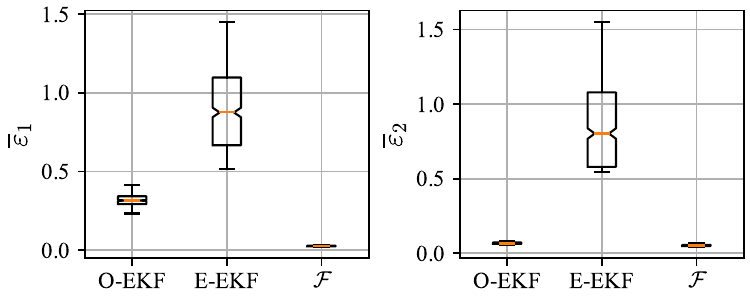}\vspace{-.2cm}
    \caption{\textcolor{black}{Absolute estimation error: a compact representation of its mean and dispersion over time and the $100$ evaporation processes considered in testing.}}
    \label{fig:boxplot}
\end{figure}
\textcolor{black}{The results reported in \figurename{~\ref{fig:comparison_errors}} show that the absolute error achieved with both O-EKF and E-EKF is larger than the one attained with the meta-filter both in mean and standard deviation, being significantly larger than the transient estimation error attained with the proposed in-context learning strategy. While the behavior after the transient can be linked to the approximations introduced by the EKF due to linearization, the behavior in the transient can be related to another hyper-parameter that must be chosen in standard Kalman filtering and is instead not required by the meta-filter, namely the initial state. For a fair comparison, in this case, we have performed such a guess relying on the steady-state information at our disposal, always setting $\hat{x}_{0|-1}$ to the nominal steady-state value in \eqref{eq:nominal_ss_values}. The superior estimation capabilities shown over time are further confirmed by the comparison provided in \figurename{~\ref{fig:boxplot}}. From these plots, it is also clear that both versions of EKF perform better in estimating the observed state (thus working as denoising filters) rather than in reconstructing the hidden one (namely $x_{1}$), differently from what happens with the meta-filter. This result is nonetheless to be expected since EKF is not impacted by how the states are excited by the considered input, which is instead essential for the meta-filter.}\\
We finally compare the computational time required to retrieve the state estimate in real time. Thanks to their recursive nature, the CPU time required by both versions of the EKF to return a state estimate is approximately of a magnitude order lower than the one needed by the meta-filter to provide (a more accurate) estimate. Note that, the computational time required by the meta-filter to reconstruct the state would considerably be reduced (with a CPU time of $0.16 \pm 0.03$~[ms] {for each time step}) if one would ideally have the whole sequence of testing input/output data (thus working in a batch mode) and fed it to the filter. This is however not realistic in filtering applications. Nonetheless, by using the practical solution proposed in Remark~\ref{rem:deployment}, \textcolor{black}{the computational time becomes comparable to that attained with the EKFs for $n_{\mathrm{ctx}} \leq 50$.}
\begin{table}[!tb]
    \centering
    \caption{\textcolor{black}{Meta-filter \emph{vs} EKFs: CPU [ms] time to return the state estimate at each time instant}}
    \setlength{\tabcolsep}{4.25pt}
    \begin{tabular}{ccc}
        \hline
         O-EKF & E-EKF & $\mathcal{F}_{\phi}$\\
        \hline
        \textbf{2.87 $\pm$ 0.53} & 3.66 $\pm$ 0.60 & 35.72 $\pm$ 25.42\\
        \hline
    \end{tabular}
    \label{tab:comparison}
\end{table}

\section{Conclusions}\label{sec:conclusions}
\textcolor{black}{In this work, we present a novel approach for model-free state estimation based on the rationale of in-context learning. The proposed state estimator does not rely on a model of the underlying system. Rather, it exploits the in-context learning capabilities of Transformer architectures. The considered learning framework allows us to shift the tuning burden from the standard weights used in Kalman filtering (i.e., the covariance matrices) to the hyper-parameter of the Transformer, thus requiring users to undertake the tuning task only once for each class of systems rather than re-calibrating the hyper-parameter for each new system's instance within the class. Our numerical results prove the effectiveness of the proposed solution in reducing the state estimation error, at the price of an intensive (yet one-shot) offline training phase.}

Future research directions include \textcolor{black}{the adaptation of this strategy for sim-to-real state estimation, its extension to multi-step-ahead estimation for predictive control purposes, and, last but not least, its experimental validation}

\bibliography{main}             

\appendix


\end{document}